\documentclass[prd,twocolumn,showpacs,preprintnumbers,superscriptaddress,floatfix,nofootinbibnofootinbib]{revtex4-1}
\usepackage{amssymb}
\usepackage{amsmath,txfonts,amsfonts}
\usepackage{graphicx}
\usepackage{dcolumn}
\usepackage{color}
\usepackage{bm,braket}
\usepackage[subfigure]{graphfig}
\usepackage{makecell}
\usepackage[colorlinks,
            citecolor=blue,
            anchorcolor=green,
            menucolor=orange,
            linkcolor=red,
            filecolor=red,
            runcolor=pink,
            urlcolor=blue,
            frenchlinks=red]{hyperref}
\pdfpagewidth=\paperwidth
\pdfpageheight=\paperheight
\allowdisplaybreaks
\begin{document}

\title{Combined analysis of the data on cross sections and spin density matrix elements for $K^*\Sigma$ photoproduction reactions}

\author{Ai-Chao Wang}
\affiliation{School of Nuclear Science and Technology, University of Chinese Academy of Sciences, Beijing 101408, China}
\affiliation{College of Science, China University of Petroleum (East China), Qingdao 266580, China}

\author{Neng-Chang Wei}
\affiliation{School of Physics, Henan Normal University, Henan 453007, China}

\author{Fei Huang}
\email[Corresponding author. Email: ]{huangfei@ucas.ac.cn}
\affiliation{School of Nuclear Science and Technology, University of Chinese Academy of Sciences, Beijing 101408, China}

\date{\today}

\begin{abstract}

In our earlier work [Phys. Rev. C \textbf{98}, 045209 (2018)], we analyzed the differential cross-section data from the CLAS Collaboration for the reactions $\gamma p \to K^{*+}\Sigma^0$ and $\gamma p \to K^{*0} \Sigma^+$ using an effective Lagrangian approach. We found that a satisfactory description of the data required the inclusion of the $s$-channel $\Delta(1905)5/2^+$ resonance, in addition to $t$-channel exchanges of $K$, $\kappa$, and $K^*$, $s$-channel contributions from nucleon ($N$) and $\Delta$, $u$-channel exchanges of $\Lambda$, $\Sigma$, and $\Sigma^*$, and a generalized contact term. In the present work, we extend our analysis to incorporate the data on spin density matrix elements from the LEPS Collaboration for the $\gamma p \to K^{*0}\Sigma^+$ reaction at photon energies $E_{\gamma} = 1.85$--$2.96$ GeV. Our goal is to impose more stringent constraints on the theoretical model and obtain a more reliable understanding of the reaction mechanisms. We obtain two fits that describe the experimental data equally well. In both fits, the $\Delta(1905)5/2^+$ resonance plays an important role. However, the contribution from $t$-channel $\kappa$ exchange is significant in one fit but negligible in the other. This finding contradicts earlier claims in the literature that the LEPS parity spin asymmetry $P_\sigma$ data support a dominant role of $\kappa$ exchange in $\gamma p \to K^{*0}\Sigma^+$. We also present predictions for $P_\sigma$ at $E_\gamma = 8.5$ GeV, which may help clarify whether $\kappa$ exchange is indeed dominant in this reaction.

\end{abstract}



\maketitle

\section{Introduction}   \label{Sec:intro}

A deep understanding of nucleon and $\Delta$ resonances ($N^\ast$'s and $\Delta^\ast$'s) is essential for gaining insight into the nonperturbative regime of quantum chromodynamics (QCD). The increasing interest in $N^\ast$'s and $\Delta^\ast$'s reflects their fundamental role in hadron physics. Historically, knowledge of these resonances has mainly come from studies of $\pi N$ scattering and photoproduction reactions involving pions and kaons. However, many resonances predicted by quark models \cite{Isgur:1977ef, Koniuk:1979vy} and lattice QCD \cite{Edwards:2011, Edwards:2013, Engel:2013, Lang:2013, Lang:2017, Kiratidis:2017, Andersen:2018} have not yet been confirmed experimentally. These so-called ``missing resonances'' are thought to couple only weakly to the $\pi N$, $K\Lambda$, and $K\Sigma$ channels, motivating the exploration of alternative production channels with stronger couplings. In this work, we focus on the $K^* \Sigma$ photoproduction reaction, which provides access to a higher mass region of $N^*$ and $\Delta^*$ due to its relatively large threshold energy, thereby complementing studies based on $\pi N$ scattering and photoproduction reactions involving pions and kaons.

Experimentally, the $K^* \Sigma$ photoproduction process has been investigated by several collaborations \cite{Hleiqawi:2005sz, Hleiqawi:2007ad, Nanova:2008kr, Hwang2012, Wei:2013}. The CLAS Collaboration at the Thomas Jefferson National Accelerator Facility (JLab) reported high-statistics differential cross-section data for the $\gamma p \to K^{*+} \Sigma^0$ reaction  in 2013 \cite{Wei:2013}. Cross-section data for the $\gamma p \to K^{*0} \Sigma^+$ reaction are available from both the CLAS Collaboration \cite{Hleiqawi:2007ad} and the CBELSA/TAPS Collaboration \cite{Nanova:2008kr}. In addition, the LEPS Collaboration provided spin density matrix elements data for $\gamma p \to K^{*0} \Sigma^+$ at photon energies between $1.85$ and $2.96$ GeV \cite{Hwang2012}.

On the theoretical side, several studies have been dedicated to the investigation of $K^* \Sigma$ photoproduction reactions \cite{Zhao:2001jw,Oh:2006in,Kim:2013,Kim:20132,Wang:2018,Ben:2023}. In our previous work \cite{Wang:2018}, a two-channel combined analysis of the cross-section data for the $\gamma p \to K^{*+} \Sigma^0$ and $\gamma p \to K^{*0} \Sigma^+$ reactions was conducted within an effective Lagrangian approach. Our results indicated that a satisfactory description of the CLAS cross-section data \cite{Wei:2013,Hleiqawi:2007ad} requires at least one $\Delta$ resonance, specifically $\Delta(1905)5/2^+$, in constructing the reaction amplitude. Similar to the case of $K^{*} \Lambda$ photoproduction \cite{Wang:2017,Kim:2014}, $t$-channel $K$ exchange was found to dominate the forward-angle differential cross sections across the energy range considered. However, a different conclusion was reached in Ref.~\cite{Oh:2006in}, where the preliminary differential cross-section data for $K^{*0} \Sigma^+$ photoproduction \cite{Hleiqawi:2005sz} were analyzed, suggesting a significant role for $t$-channel $\kappa$ exchange in this reaction. Additionally, in Ref.~\cite{Hwang2012}, the LEPS parity spin asymmetry $P_{\sigma}$ data were interpreted as supporting the model II in Ref.~\cite{Oh:2006in} that includes a substantial $\kappa$-exchange contribution. More recently, Ref.~\cite{Ben:2023} proposed the $N(2080)3/2^-$ and $N(2270)3/2^-$ molecular states as alternatives to the $\Delta(1905)5/2^+$ resonance and found that the differential cross-section data could still be reproduced.

Despite these efforts, theoretical investigations of $K^*\Sigma$ photoproduction have largely concentrated on differential cross-section data \cite{Hleiqawi:2005sz, Hleiqawi:2007ad, Nanova:2008kr, Wei:2013}. A comprehensive analysis that simultaneously accounts for data on both differential cross sections and spin density matrix elements has been lacking.

In the present work, we extend our earlier study \cite{Wang:2018} by incorporating the LEPS data on spin density matrix elements for $\gamma p\to K^{*0}\Sigma^+$ \cite{Hwang2012} into the analysis. Our goal is to better constrain the theoretical model through a simultaneous description of the available data on both differential cross sections and spin density matrix elements for both $\gamma p\to K^{*+}\Sigma^0$ and $\gamma p\to K^{*0}\Sigma^+$ reactions. Such an approach is expected to yield a more refined understanding of the underlying reaction mechanisms, particularly the role of resonance contributions and the possible involvement of $\kappa$-meson exchange.

The paper is organized as follows. Section~\ref{Sec:formalism} outlines the theoretical framework. Section~\ref{Sec:results} presents the numerical results and discusses the mechanisms of the reactions considered. A brief summary and conclusions are given in Sec.~\ref{sec:summary}.

\section{Formalism}  \label{Sec:formalism}

\begin{figure}[tbp]
\centering
{\vglue 0.15cm}
\subfigure[~$s$ channel]{
\includegraphics[width=0.45\columnwidth]{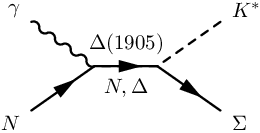}}  {\hglue 0.4cm}
\subfigure[~$t$ channel]{
\includegraphics[width=0.45\columnwidth]{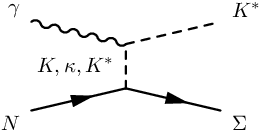}} \\[6pt]
\subfigure[~$u$ channel]{
\includegraphics[width=0.45\columnwidth]{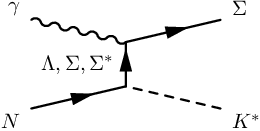}} {\hglue 0.4cm}
\subfigure[~Interaction current]{
\includegraphics[width=0.45\columnwidth]{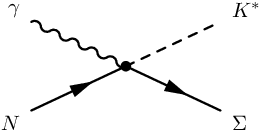}}
\caption{Generic structure of photoproduction amplitude for $\gamma N\to K^{*}\Sigma$. Time proceeds from left to right.}
\label{FIG:feymans}
\end{figure}

\begin{table}[tbp]
\caption{\label{Table:para} Values of adjustable parameters. $A_{1/2}$ and $A_{3/2}$ represent helicity amplitudes for the resonance. The asterisks ($\ast$$\ast$$\ast$$\ast$) denote the overall status of the resonance evaluated by PDG \cite{PDG:2024}. The numbers in brackets below the resonance masses, widths, and helicity amplitudes represent the corresponding values estimated by PDG \cite{PDG:2024}.}
\begin{tabular*}{\columnwidth}{@{\extracolsep\fill}lcr}
\hline\hline
                     & Model I & Model II \\ \hline
$g^{(2)}_{\Sigma^{*+} \Sigma^{+} \gamma}/g^{(1)}_{\Sigma^{*+} \Sigma^{+} \gamma}$ & $ -1.83\pm 0.76$&$0.50 \pm 0.74 $ \\
$g^{(1)}_{\Sigma^{*0} \Sigma^{0} \gamma}$  & $ -2.54\pm0.68 $&$ -0.00029\pm 0.0003$ \\
$g^{(2)}_{\Sigma^{*0} \Sigma^{0} \gamma}$  & $ 2.62\pm 0.36$&$ -5.78\pm 2.8 $ \\
$g_{\Delta \Sigma K^*}^{(1)}$ & $ 20.14\pm 2$&$ -25.0\pm 3.6$ \\
$\Lambda_N$ [MeV]   & $ 1737\pm 28$&$ 1000\pm 15$ \\
$\Lambda_\Delta$ [MeV]   & $ 1368\pm 10$&$ 1137\pm 24 $ \\
$\Lambda_{\Sigma^*}$  [MeV]  & $ 857\pm 22$&$ 838\pm 18$ \\
$\Lambda_{\Lambda}$ [MeV]  & $ 745\pm 25$&$ 779\pm15 $ \\
$\Lambda_{\Sigma}$ [MeV]  & $ 1194\pm 9$&$ 1173\pm 9$ \\
$\Lambda_{K}$  [MeV]  & $ 700\pm 10$&$ 833\pm 19$ \\
$\Lambda_{\kappa}$  [MeV]  & $ 1000\pm 2$&$ 1800\pm 2$ \\
$\Lambda_{K^{*}}$  [MeV]  & $ 1100\pm 3$&$ 1293\pm 15$ \\
\hline
$\Delta(1905){5/2}^+$ ($\ast$$\ast$$\ast$$\ast$)  &    &  \\
$M_R$ [MeV] &  $1855\pm 2$ &$ 1910\pm 0.4$    \\
	             &  $[1855 \sim 1910]$ & $[1855 \sim 1910]$ \\
$\Gamma_R$ [MeV]  & $ 346\pm 91$ &$ 400\pm 2$   \\
					&   $[270\sim 400]$ & $[270\sim 400]$ \\
$A_{1/2}$ [GeV$^{-1/2}$] & $0.025 \pm0.007 $&$ 0.017\pm0.0001 $ \\
                    &   $[0.017\sim 0.027]$ & $[0.017\sim 0.027]$ \\
$A_{3/2}$ [GeV$^{-1/2}$] & $-0.038 \pm0.013 $&$ -0.055\pm 0.0002$ \\
                    &   $[-0.055\sim -0.035]$ & $[-0.055\sim -0.035]$ \\
$g^{(1)}_{R \Sigma K^*}$ & $ 11.52\pm 0.20$&$ 0.316\pm0.005 $ \\
$g^{(2)}_{R \Sigma K^* }$ & $ -0.137\pm 0.01$&$ -5.78\pm 0.007$ \\
$g^{(3)}_{R \Sigma K^* }$ & $ 115.8\pm 1.18$&$ 43.5\pm 0.001$ \\
$\Lambda_R$ [MeV] & $ 1150\pm 9$ &$1307 \pm 9$   \\
\hline\hline
\end{tabular*}
\end{table}

\begin{figure*}[tbp]
\includegraphics[width=0.75\textwidth]{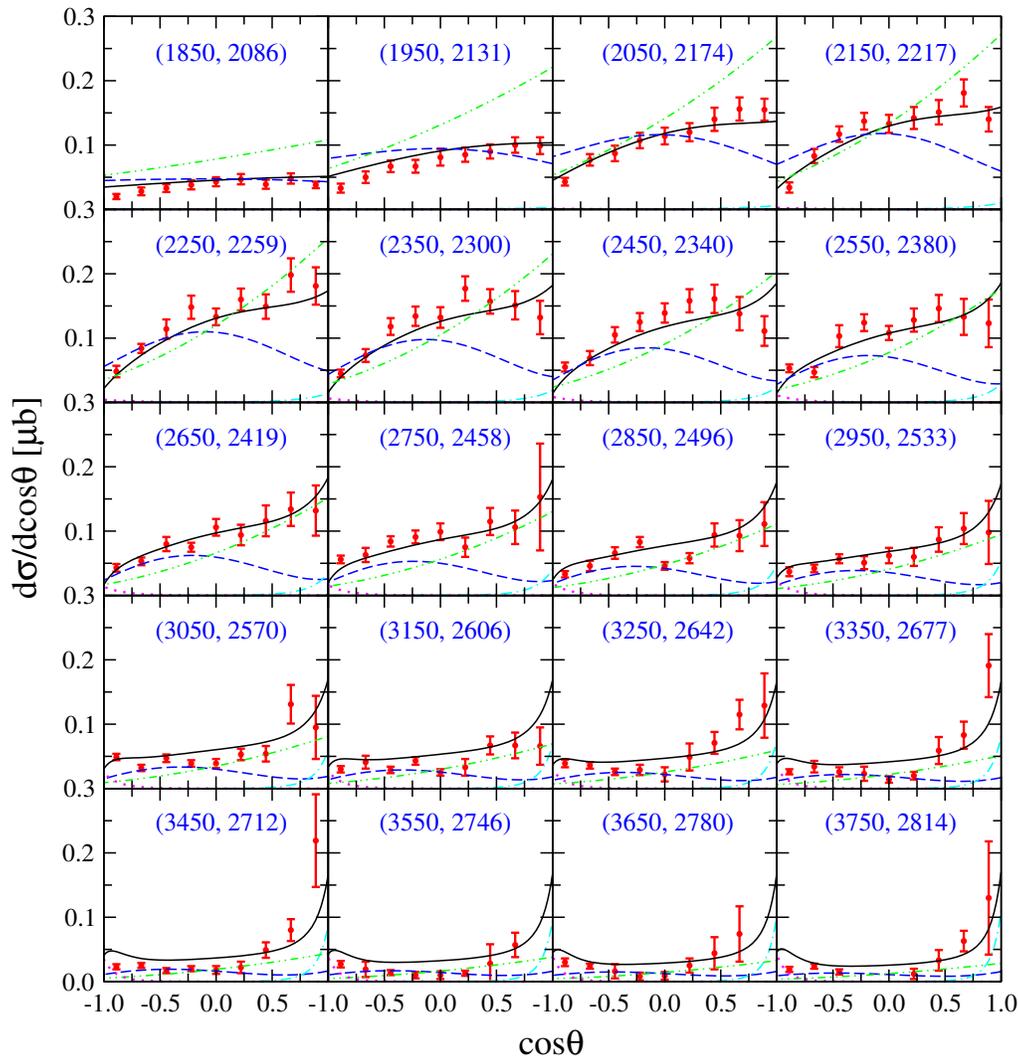}
\caption{Differential cross sections for $\gamma p \to K^{*+}\Sigma^0$ as a function of $\cos\theta$ (black solid lines) in model I. The scattered symbols denote the CLAS data in Ref.~\cite{Wei:2013}. The blue dashed, green dash-double-dotted, magenta dotted, and cyan dash-dotted lines represent the individual contributions from the $\Delta(1905)5/2^+$, $\Delta$, $\Lambda$, and $K^*$ exchanges, respectively. The numbers in parentheses denote the centroid value of the photon laboratory incident energy (left number) and the corresponding total c.m. energy of the system (right number), in MeV.}
\label{fig:dif1}
\end{figure*}

\begin{figure*}[tbp]
\includegraphics[width=0.75\textwidth]{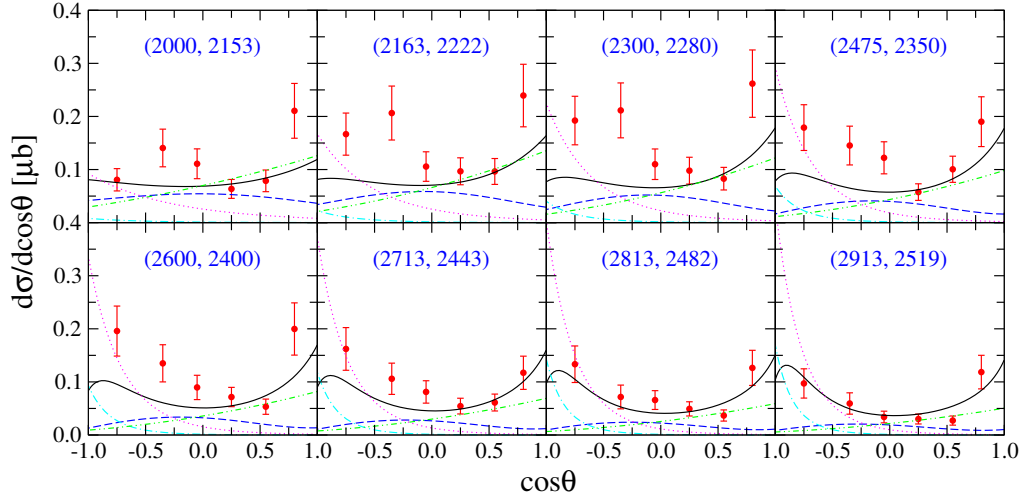}
\caption{Differential cross sections for $\gamma p \to K^{*0}\Sigma^+$ as a function of $\cos\theta$ in model I. Notations are the same as in Fig.~\ref{fig:dif1} except that now the cyan dash-dotted lines represent the contributions from the $u$-channel $\Sigma^*$ exchange, the magenta dotted lines denote the contributions from the $u$-channel $\Sigma$ exchange, and the scattered symbols denote the CLAS data in Ref.~\cite{Hleiqawi:2007ad}.}
\label{fig:dif2}
\end{figure*}

\begin{figure*}[tbp]
\includegraphics[width=0.75\textwidth]{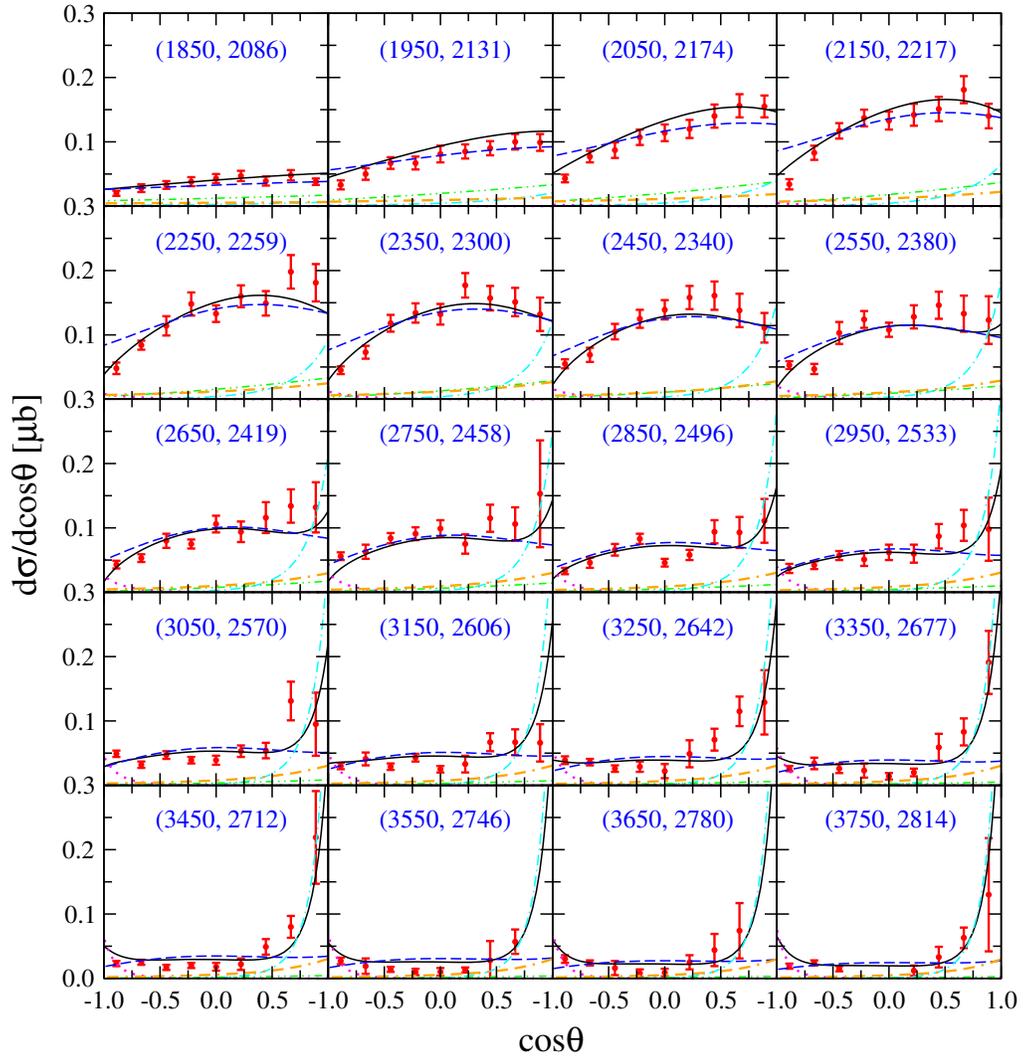}
\caption{Differential cross sections for $\gamma p \to K^{*+}\Sigma^0$ as a function of $\cos\theta$ in model II. Notations are the same as in Fig.~\ref{fig:dif1} except that the orange dot-double-dashed lines represent the contributions from the $\kappa$ exchange.}
\label{fig:dif1_2}
\end{figure*}

\begin{figure*}[tbp]
\includegraphics[width=0.75\textwidth]{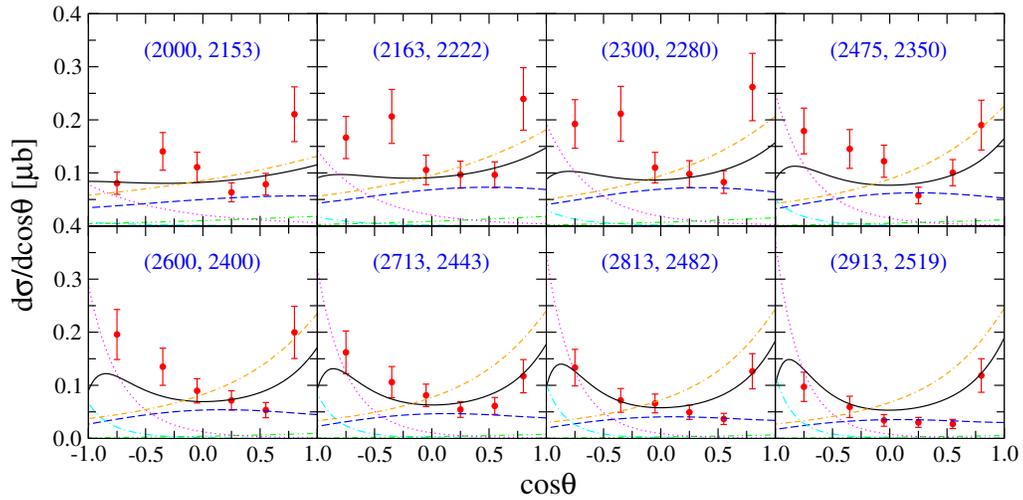}
\caption{Differential cross sections for $\gamma p \to K^{*0}\Sigma^+$ as a function of $\cos\theta$ in model II. Notations are the same as in Fig.~\ref{fig:dif2} except that the orange dot-double-dashed lines represent the contributions from the $\kappa$ exchange.}
\label{fig:dif2_2}
\end{figure*}

\begin{figure}[tbp]
\includegraphics[width=0.4\textwidth]{sig1} \\[9pt]
\includegraphics[width=0.4\textwidth]{sig2}
\caption{ Total cross sections with dominant individual contributions in model I. The scattered symbols are data from CLAS Collaboration \cite{Wei:2013}. Notations in upper and lower graphs are the same as in Figs.~\ref{fig:dif1} and \ref{fig:dif2}, respectively.}
\label{fig:total_cro_sec}
\end{figure}

\begin{figure}[tbp]
\includegraphics[width=0.4\textwidth]{sig1_2} \\[9pt]
\includegraphics[width=0.4\textwidth]{sig2_2}
\caption{Total cross sections with dominant individual contributions in model II. The scattered symbols are data from CLAS Collaboration \cite{Wei:2013}. Notations in upper and lower graphs are the same as in Figs.~\ref{fig:dif1_2} and \ref{fig:dif2_2}, respectively.}
\label{fig:total_cro_sec_2}
\end{figure}

\begin{figure*}[tbp]
\includegraphics[width=0.75\textwidth]{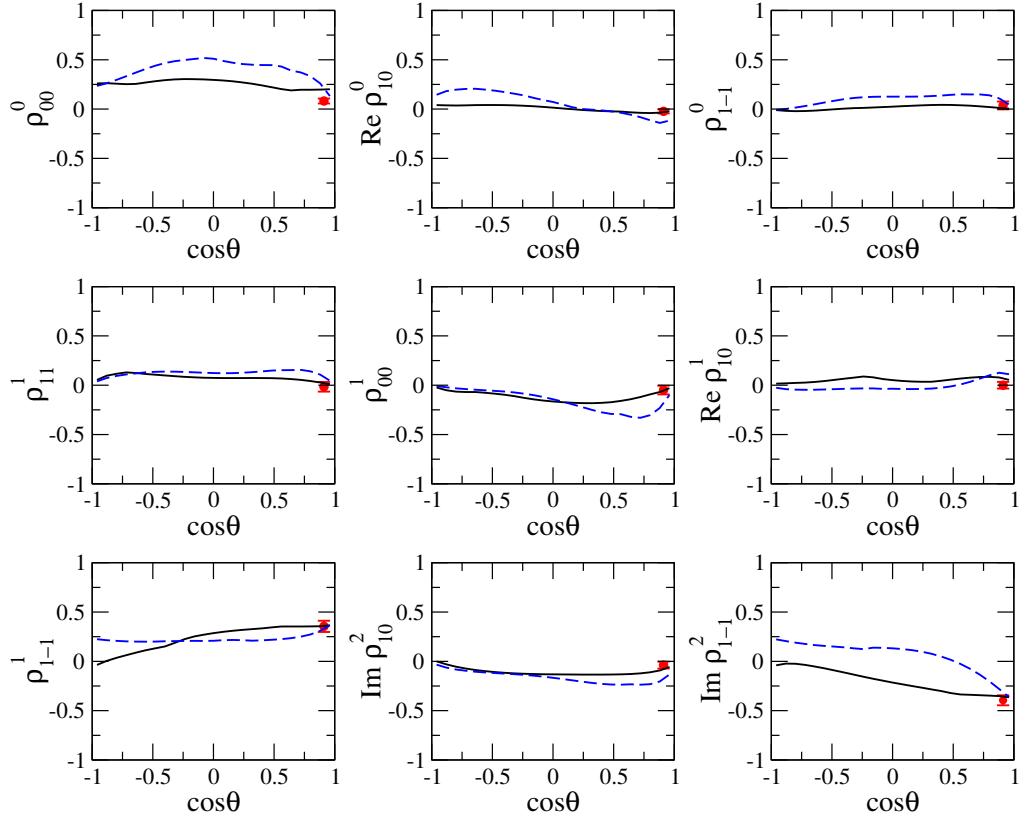}
\caption{Spin density matrix elements for $\gamma p \to K^{*0}\Sigma^+$ as a function of $\cos\theta$ at $E_{\gamma}=1.85$--$2.96$ GeV. The black solid line and blue dashed line correspond to the results in model I and model II, respectively. The data are from the LEPS Collaboration \cite{Hwang2012}. }
\label{fig:sdme}
\end{figure*}

\begin{figure}[tbp]
\includegraphics[width=0.4\textwidth]{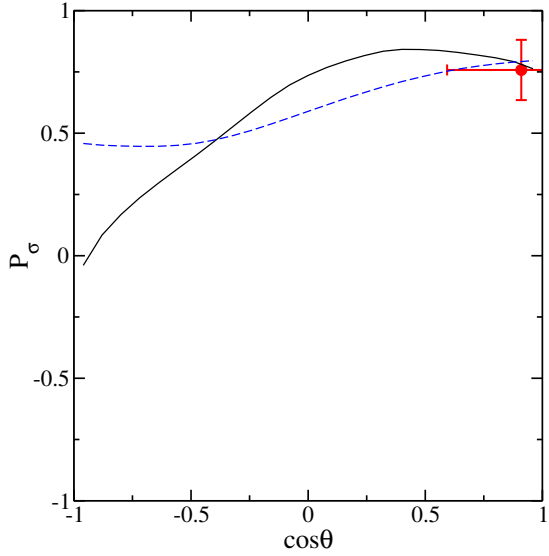}
\caption{Parity spin asymmetry ($P_{\sigma}=2\rho^{1}_{1-1}-\rho^{1}_{00}$) for $\gamma p \to K^{*0}\Sigma^+$ as a function of $\cos\theta$ at $E_{\gamma}=1.85$--$2.96$ GeV. The black solid line and blue dashed line correspond to the results in model I and model II, respectively. The data are from the LEPS Collaboration \cite{Hwang2012}. }
\label{fig:psig}
\end{figure}

\begin{figure}[tbp]
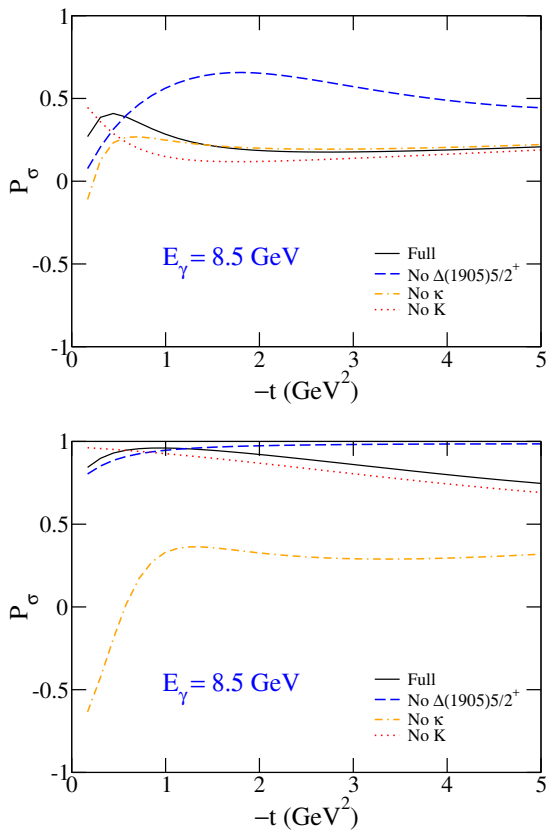

\includegraphics[width=0.4\textwidth]{40_sdme} \\[9pt]
\includegraphics[width=0.4\textwidth]{40_sdme_2}
\caption{Parity spin asymmetry for $\gamma p \to K^{*0}\Sigma^+$ as a function of $-t$ at $E_{\gamma}=8.5$ GeV. The upper and lower graphs correspond results in model I and model II, respectively.}
\label{fig:psig_85}
\end{figure}

Following the formalism developed in Ref.~\cite{Wang:2018}, the scattering amplitudes for $\gamma N \to K^* \Sigma$ in the present work are constructed by considering the $t$-channel exchanges of $K$, $\kappa$, and $K^*$ mesons, $s$-channel contributions of $N$, $\Delta$, and $\Delta(1905)5/2^+$ baryons, $u$-channel exchanges of $\Lambda$, $\Sigma$, and $\Sigma^*$ baryons, and a generalized contact term, as schematically depicted in Fig.~\ref{FIG:feymans}. The full reaction amplitude can be expressed according to a full field theoretical approach of Refs.~\cite{Haberzettl:1997,Haberzettl:2006,Huang:2012,Huang:2013} as
\begin{eqnarray}
M^{\nu\mu} = M^{\nu\mu}_s + M^{\nu\mu}_t + M^{\nu\mu}_u + M^{\nu\mu}_{\rm int},  \label{eq:amplitude}
\end{eqnarray}
where $\nu$ and $\mu$ are the Lorentz indices of vector meson $K^*$ and photon $\gamma$, respectively; $M^{\nu\mu}_s$, $M^{\nu\mu}_t$, and $M^{\nu\mu}_u$ stand for the $s$-, $t$-, and $u$-channel amplitudes, respectively; and $M^{\nu\mu}_{\rm int}$ is a generalized interaction current. The amplitudes $M^{\nu\mu}_s$, $M^{\nu\mu}_t$, and $M^{\nu\mu}_u$ can be obtained by direct evaluations of the corresponding Feynman diagrams. The interaction current $M^{\nu\mu}_{\rm int}$ arises from the photon attaching inside the $\Sigma N K^*$ interaction vertex. It is in principle highly nonlinear and involves very complicated diagrams, making a strict calculation impractical. We follow Refs.~\cite{Haberzettl:1997,Haberzettl:2006,Huang:2012,Huang:2013} to choose a particular prescription for $M^{\nu\mu}_{\rm int}$, which obeys the crossing symmetry and ensures that the full photoproduction amplitude for $\gamma N \to K^* \Sigma$ satisfies the generalized Ward-Takahashi identity and thus is fully gauge invariant. The explicit expressions of the Lagrangians, propagators, and form factors needed for calculating the amplitudes $M^{\nu\mu}_s$, $M^{\nu\mu}_t$, and $M^{\nu\mu}_u$ and the adopted prescription for $M^{\nu\mu}_{\rm int}$ can be found in our previous work \cite{Wang:2018}, and we do not repeat them here.

In the center-of-mass (c.m.) frame, the $\gamma p \to K^{*0} \Sigma^{+}$ invariant amplitude $M^{\nu\mu}$ introduced in Eq.~(\ref{eq:amplitude}) can be expressed in helicity basis as \cite{Jacob:1964}
\begin{equation}
T_{\lambda_V\lambda_f,\lambda_\gamma\lambda_i} \! \left(W,\theta\right) \equiv \braket{q,\lambda_V; p_f, \lambda_f | M | k, \lambda_\gamma; p_i, \lambda_i},     \label{eq:HelMtrx}
\end{equation}
where $\lambda_i$, $\lambda_\gamma$, $\lambda_f$, and $\lambda_V$ denote the helicities of the incoming nucleon, incoming photon, outgoing $\Sigma$, and outgoing $K^\ast$, respectively. Correspondingly, the arguments ${ p}_i$, ${ k}$, ${ p}_f$, and ${ q}$ represent the 4-momenta of the incoming nucleon, incoming photon, outgoing $\Sigma$, and outgoing $K^*$, respectively. $W$ and $\theta$ indicate the total energy of the system and the scattering angle in c.m. frame, respectively. The differential cross section is then given by
\begin{equation}
\frac{d\sigma}{d\Omega} = \frac{1}{64\pi^2 W^2} \frac{|\bm{q}|}{|\bm{k}|} \frac{1}{4} \sum_{\lambda_V,\lambda_f,\lambda_\gamma,\lambda_i} \left | T_{\lambda_V\lambda_f,\lambda_\gamma\lambda_i} \!\left(W,\theta\right)\right|^2,   \label{eq:3-1}
\end{equation}
and the spin density matrix elements relevant to the present work are given by \cite{Wei:2019,Schilling:1969um}
\begin{align}
\rho^0_{\lambda_V \lambda'_V}  &= \frac{\sum_{\lambda_f,\, \lambda_\gamma ,\, \lambda_i} T_{\lambda_V\lambda_f, \lambda_\gamma\lambda_i} {T^*_{\lambda'_V\lambda_f,\lambda_\gamma\lambda_i}} }{\sum_{\lambda_V,\,\lambda_f ,\,\lambda_\gamma ,\,\lambda_i} \left | T_{\lambda_V\lambda_f,\lambda_\gamma\lambda_i}\right|^2},   \\[6pt]
\rho^1_{\lambda_V \lambda'_V}   &= \frac{\sum_{\lambda_f ,\,\lambda_\gamma,\, \lambda_i} T_{\lambda_V\lambda_f, -\lambda_\gamma\lambda_i} {T^*_{\lambda'_V\lambda_f,\lambda_\gamma\lambda_i}} }{\sum_{\lambda_V,\,\lambda_f,\, \lambda_\gamma ,\,\lambda_i} \left | T_{\lambda_V\lambda_f,\lambda_\gamma\lambda_i} \right|^2},   \\[6pt]
\rho^2_{\lambda_V \lambda'_V}   &= i \,\frac{\sum_{\lambda_f ,\,\lambda_\gamma,\, \lambda_i}\lambda_\gamma  T_{\lambda_V\lambda_f, -\lambda_\gamma\lambda_i} {T^*_{\lambda'_V\lambda_f,\lambda_\gamma\lambda_i}} }{\sum_{\lambda_V,\,\lambda_f,\, \lambda_\gamma,\, \lambda_i} \left | T_{\lambda_V\lambda_f,\lambda_\gamma\lambda_i} \right|^2}.
\label{eq:SDMEs}
\end{align}

\section{Results and discussion}   \label{Sec:results}

In our previous work \cite{Wang:2018}, we conducted a two-channel combined analysis of the CLAS cross-section data \cite{Wei:2013, Hleiqawi:2007ad} for the $\gamma p \to K^{*+} \Sigma^0$ and $\gamma p \to K^{*0} \Sigma^+$ reactions using an effective Lagrangian approach. By considering the $t$-channel $K$, $\kappa$, and $K^*$ exchanges, the $s$-channel $N$, $\Delta$, and $\Delta(1905)5/2^+$ contributions, the $u$-channel $\Lambda$, $\Sigma$, and $\Sigma^*$ exchanges, and a generalized contact term, the CLAS cross-section data were satisfactorily described.
In the present work, we extend this analysis to incorporate the data on spin density matrix elements from the LEPS Collaboration for the $\gamma p\to K^{*0}\Sigma^+$ reaction at photon energies between $1.85$ and $2.96$ GeV \cite{Hwang2012}. The main purpose is to impose more stringent restrictions on our theoretical model. Note that in the literature, the LEPS data on spin density matrix elements has never been analyzed by theoretical works. Here, we for the first time perform a combined analysis of the available data on both cross sections and spin density matrix elements for both the $\gamma p \to K^{*+} \Sigma^0$ and $\gamma p \to K^{*0} \Sigma^+$ reactions. It is expected that a more reliable understanding of the reaction mechanisms, in particular, the role of resonance contributions, could be achieved. Moreover, we try to clarify through such a combined analysis whether the $\kappa$-meson exchange is dominant or not in $\gamma p \to K^{*0} \Sigma^+$ as claimed in literature \cite{Oh:2006in,Hwang2012}.

Using the parameters from Ref.~\cite{Wang:2018} that were fixed by fitting only the cross-section data, we found that the LEPS spin density matrix element data for $\gamma p \to K^{*0} \Sigma^+$ could not be satisfactorily reproduced, as significant discrepancies between theoretical predictions and the data are evident. We therefore treat the cutoff masses for the $s$-channel $N$, $\Delta$, and $\Delta(1905)5/2^+$ contributions--$\Lambda_N$, $\Lambda_\Delta$, and $\Lambda_{\Delta(1905)}$--which were set to a common parameter $\Lambda_s$ in Ref.~\cite{Wang:2018}, as independent free parameters. Similarly, we treat the cutoff masses for the $t$-channel $K$ and $\kappa$ exchanges--$\Lambda_K$ and $\Lambda_\kappa$--which were set to a common parameter $\Lambda_{K,\kappa}$ in Ref.~\cite{Wang:2018}, as independent free parameters. We then refit all available data, including the newly considered spin density matrix element data for the $K^{*0}\Sigma^+$ reaction channel and the previously considered differential cross-section data for both $K^{*+}\Sigma^0$ and $K^{*0}\Sigma^+$ reaction channels. After numerous attempts with various starting values for the fitting parameters, we obtained two sets of parameters that effectively reproduce all available data. Both fits perform slightly worse in terms of the differential cross-section data compared to the fit in Ref.~\cite{Wang:2018} (where $\chi^2_{{\rm d}\sigma}/{\rm ND}_{{\rm d}\sigma} = 1.84$), yielding  $\chi^2_{{\rm d}\sigma}/{\rm ND}_{{\rm d}\sigma} = 1.97$ and $2.02$, respectively. This decrease in fit quality is expected, because only differential cross-section data were considered in Ref.~\cite{Wang:2018}, whereas the addition of spin density matrix element data inevitably lowers the quality of the fit to the differential cross-section data. It is worth noting that models I and II have identical resonance content and the same number of free parameters. Therefore, statistical criteria such as AIC or BIC reduce effectively to a comparison of $\chi^2$, for which the two models yield very similar values. The existence of two solutions with comparable fit quality reflects a physical ambiguity arising from the limited constraints provided by the currently available data. Additional experimental data, especially data at higher energies on polarization observables, are needed to resolve this ambiguity.

In Table~\ref{Table:para}, we present the fitting values of all the adjustable parameters. The two  fits are named as model I and model II, respectively. Here, the uncertainties in the resulting parameters arise from the uncertainties (error bars) associated with the fitted data points. For $\gamma p\to K^{*0}\Sigma^+$, the PDG \cite{PDG:2024} value of the partial decay width $\Gamma_{\Sigma^{*+} \to \Sigma^{+} \gamma}$ is used to constrain the electromagnetic couplings $g^{(1)}_{\Sigma^{*+} \Sigma^{+} \gamma}$ and $g^{(2)}_{\Sigma^{*+} \Sigma^{+} \gamma}$, and we treat their ratio $g^{(2)}_{\Sigma^{*+} \Sigma^{+} \gamma}/g^{(1)}_{\Sigma^{*+} \Sigma^{+} \gamma}$ as a fitting parameter in practice. For $\gamma p \to K^{*+}\Sigma^0$, as no electromagnetic decay information for $\Sigma^{*0}$ is available, both $g^{(1)}_{\Sigma^{*0} \Sigma^{0} \gamma}$ and $g^{(2)}_{\Sigma^{*0} \Sigma^{0} \gamma}$ are treated as fitting parameters.

Focusing on the cutoff masses of the non-resonant terms in the two models, we observe that most of the fitting values are very close, except for the cutoffs of the $s$-channel $N$ exchange and the $t$-channel $\kappa$ exchange. The relatively large difference in the cutoffs of $N$ exchange has little effect on the models, because the contributions from $N$ exchange are minimal in both models. However, for the $t$-channel $\kappa$ exchange, there is a significant difference: with a cutoff value of $1800$ MeV in model II, the $\kappa$ exchange contributes significantly, whereas with a cutoff value of $1000$ MeV in model I, its contribution is nearly negligible. This constitutes a fundamental difference between the two models. $M_R$, $\Gamma_R$, and $\Lambda_R$ denote the resonance mass, width, and cutoff mass, respectively. The asterisks ($\ast$$\ast$$\ast$$\ast$) following the resonance name indicate the overall status of the resonance as evaluated in the most recent review by the PDG \cite{PDG:2024}. The numbers in brackets below the resonance mass, width, and helicity amplitudes are the corresponding values estimated by the PDG. $A_{1/2}$ and $A_{3/2}$ stand for the helicity amplitudes for the resonance radiative decay to $\gamma p$. For the four-star resonance $\Delta(1905)5/2^+$, in Ref.~\cite{Wang:2018}, we used the averaged values from the PDG for its mass, width, and helicity amplitudes. However, for the two models in this work, we fit the parameters within the recommended value ranges from the PDG for these three parameters. The fitted mass of the $\Delta(1905)5/2^+$ resonance is found to lie at the boundary of the PDG range, indicating that the present data do not strongly constrain this parameter. We have examined the effect of releasing the PDG constraint on its mass and find that the fitted value can shift away from the PDG interval, with only a marginal improvement in the description of the data. This suggests that the available data are not sufficiently sensitive to determine the mass of $\Delta(1905)5/2^+$ precisely.

The $\Delta(1905)5/2^+$ resonance is known to predominantly decay into non-strange channels according to the PDG, and no explicit information on its decay into the $K^*\Sigma$ channel is currently available. Note that the threshold for $K^*\Sigma$ production is about $2082$ MeV, which is significantly higher than the mass of the $\Delta(1905)5/2^+$ resonance. Therefore, the contribution of $\Delta(1905)5/2^+$ to the present reaction arises mainly from off-shell effects. This is the reason why its on-shell decay properties listed by the PDG contain no information on the $K^*\Sigma$ channel. The importance of the $\Delta(1905)5/2^+$ resonance in the present analysis should be understood as a dynamical effect, rather than as evidence for a large decay branching ratio.

The uncertainties of the fitted parameters listed in Table~\ref{Table:para} are obtained from the covariance matrix provided by the MINUIT minimization procedure. These uncertainties arise from the propagation of the experimental errors (error bars) of the fitted data points into the parameter space. These uncertainties reflect the statistical errors propagated from the experimental data.

The results for the differential cross-sections of $\gamma p\to K^{*+}\Sigma^0$ and $\gamma p\to K^{*0}\Sigma^+$ corresponding to the parameters of model I listed in Table~\ref{Table:para} are shown in Figs.~\ref{fig:dif1} and \ref{fig:dif2}, respectively. In these figures, the black solid lines represent the full results. The blue dashed and green dash-double-dotted lines represent the individual contributions from the $\Delta(1905)5/2^+$ and $\Delta$ exchanges, respectively. The cyan dash-dotted lines represent the individual contributions from the $K^*$ exchange for $\gamma p\to K^{*+}\Sigma^0$ and the $\Sigma^*$ exchange for $\gamma p\to K^{*0}\Sigma^+$. The magenta dotted lines represent the individual contributions from the $\Lambda$ exchange for $\gamma p\to K^{*+}\Sigma^0$ and the $\Sigma$ exchange for $\gamma p\to K^{*0}\Sigma^+$. Contributions from other terms are too small to be clearly seen at the scale used and are therefore not plotted. The numbers in parentheses denote the centroid value of the photon laboratory incident energy (left number) and the corresponding total c.m. energy of the system (right number), in MeV. As stated in Ref.~\cite{Wang:2018}, the statistical data binning for photon incident energy is $100$ MeV, and its effects for $\gamma p\to K^{*+}\Sigma^0$ at the c.m. energy $W=2086$ MeV, which is about $2$ MeV higher than the $K^{*+}\Sigma^0$ threshold, have been approximated by an integral of the differential cross-sections over the $100$ MeV energy bin. At other energies, the binning effects have been tested to be minimal.

Similar to model I, the results for the differential cross-sections of $\gamma p\to K^{*+}\Sigma^0$ and $\gamma p\to K^{*0}\Sigma^+$ corresponding to the parameters of model II are shown in Figs.~\ref{fig:dif1_2} and \ref{fig:dif2_2}, respectively. In addition to the annotations as in Figs.~\ref{fig:dif1} and \ref{fig:dif2}, the orange dot-double-dashed lines represent the contributions from the $\kappa$ exchange.

From Figs.~\ref{fig:dif1}--\ref{fig:dif2} for model I and Figs.~\ref{fig:dif1_2}--\ref{fig:dif2_2} for model II, it is evident that both models provide satisfactory overall descriptions of the differential cross-section data. The fitting qualities of the two models are nearly identical, and the resulted values of chi-square per number of data, $\chi^2_{{\rm d}\sigma}/{\rm ND}_{{\rm d}\sigma}$, are $1.97$ for model I and $2.02$ for model II. However, a completely different reaction mechanism exists in these two models, where the roles played by individual terms differ significantly between them.

For the $\gamma p\to K^{*+}\Sigma^0$ reaction process, in model I, the $\Delta(1905)5/2^+$ resonance and the $\Delta$ exchange jointly contribute to the angular distribution structure near the threshold energy region, with both making substantial contributions. In contrast, model II exhibits a different behavior: only the $\Delta(1905)5/2^+$ resonance contributes dominantly near the threshold energy region, while the $\Delta$ exchange contributes much less than in model I. The $u$-channel $\Lambda$ exchange in both models contributes minimally at backward angles. The $t$-channel $K^*$ exchange is responsible for the rapid increase at forward angles in the higher energy region in model I, while in model II, in addition to the $K^*$ exchange, the $\kappa$ exchange also contributes considerably at forward angles.

Turning to the $\gamma p\to K^{*0}\Sigma^+$ reaction process, the dominant terms at each scattering angle are not as clear-cut as in the $\gamma p\to K^{*+}\Sigma^0$ process, with many terms contributing significantly over the entire energy region considered. For the $\Delta(1905)5/2^+$ resonance and $\Delta$ exchange, which have identical electromagnetic coupling constants in both reaction channels, their contributions differ in the two models due to their isospin factors. In both models, the $u$-channel $\Sigma$ exchange, followed by the $\Sigma^*$ exchange, has considerable contributions at backward angles in the considered energy region, with the contributions being nearly the same in both models due to the fitted values of the cutoff masses of these two terms being very close, as shown in Table~\ref{Table:para}. In model I, the $\Delta$ exchange mainly accounts for the increases of differential cross sections at forward angles, while in model II, the $\Delta(1905)5/2^+$ resonance interacts with the $\kappa$ exchange to produce these increases.

It is important to note that for the $K^{*0}\Sigma^+$ channel, there is almost no contribution from the $\kappa$ exchange in model I, while the $\kappa$ exchange contributes significantly, especially at the forward angles, in model II. As mentioned previously, one of the motivations of this study is to determine the contributing term at forward angles in the $K^{*0}\Sigma^+$ reaction channel. Clearly, with the spin density matrix elements data well reproduced which will be discussed in detail later, the results in model I challenge the claim in Ref.~\cite{Hwang2012} that the $\kappa$ exchange play a significant role in the $\gamma p\to K^{*0}\Sigma^+$ reaction process.

Figures~\ref{fig:total_cro_sec} (model I) and \ref{fig:total_cro_sec_2} (model II) depict the predicted total cross sections (black solid lines) along with individual contributions from the $\Delta(1905)5/2^+$ exchange (blue dashed lines), the $\Delta$ exchange (green dash-double-dotted lines), and the $\kappa$ exchange (orange dot-double-dashed lines) for the $\gamma p\to K^{*+}\Sigma^0$ (upper graph) and $\gamma p\to K^{*0}\Sigma^+$ (lower graph) reaction processes. The cyan dash-dotted lines represent the $K^*$ exchange in the upper graph and the $\Sigma^*$ exchange in the lower one. The magenta dotted lines represent the $\Lambda$ exchange in the upper graph and the $\Sigma$ exchange in the lower one. Contributions from other terms are not plotted due to their negligible impact at the scale used. The total cross sections are obtained by integrating corresponding differential cross sections over $\cos\theta$ and are not included in our fitting process.

It is evident that for the $\gamma p\to K^{*+}\Sigma^0$ reaction process (upper graphs in Figs.~\ref{fig:total_cro_sec}--\ref{fig:total_cro_sec_2}), in model I, the $\Delta(1905)5/2^+$ resonance, together with the $\Delta$ exchange, is responsible for the bump structure exhibited by the total cross-section data, while in model II, the $\Delta$ exchange contributes much less than the $\Delta(1905)5/2^+$ resonance. For the $\gamma p\to K^{*0}\Sigma^+$ reaction process (lower graphs in Figs.~\ref{fig:total_cro_sec}--\ref{fig:total_cro_sec_2}), in addition to the $\Delta(1905)5/2^+$ resonance and the $\Sigma$ exchange, which play significant roles in both models, the $u$-channel $\Sigma^*$ exchange also makes considerable contributions. The distinguishing features of these two models arise from the contributions of $\Delta$ and $\kappa$ exchanges. Specifically, in model I the $\Delta$ exchange provides rather significant contributions and the contributions from $\kappa$ exchange is negligible, while in model II, the $\kappa$ exchange plays a dominant role and the contributions from $\Delta$ exchange is rather small.

Figure~\ref{fig:sdme} shows the results on spin density matrix elements compared with the LEPS data \cite{Hwang2012}. It is seen that the data are well reproduced in both model I and model II. Note that these data were extracted using an unbinned extended maximum likelihood fit in the helicity frame, covering the beam energy region from $1.85$ to $2.96$ GeV. In practice, our theoretical results are calculated at the centroid value over this $1.11$ GeV energy bin.

Figure~\ref{fig:psig} demonstrates the results on parity spin asymmetry $P_{\sigma}$ for the $\gamma p\to K^{*0}\Sigma^+$ reaction. The observable $P_{\sigma}$ is related to spin density matrix elements $\rho^{1}_{1-1}$ and $\rho^{1}_{00}$ through $P_{\sigma}=2\rho^{1}_{1-1}-\rho^{1}_{00}$. It is seen that the LEPS $P_{\sigma}$ data \cite{Hwang2012} is also well reproduced in both model I and model II. In literature, it is believed that the value of $P_{\sigma}$ data being close to unity indicates a dominant contribution coming from natural parity exchanges. In particular, as the LEPS data on $P_{\sigma}$ for $\gamma p\to K^{*0}\Sigma^+$ tends to $1$, it is claimed in Ref.~\cite{Hwang2012} that the $t$-channel $\kappa$ exchange plays a dominant role in this reaction. In our present calculations, the $\kappa$ exchange provides almost no contributions in model I, but significant contributions in model II. However, both models describe the $P_{\sigma}$ data equally well. This means that the LEPS $P_\sigma$ data for $\gamma p\to K^{*0}\Sigma^+$ \cite{Hwang2012} does not necessarily support a dominant $\kappa$ exchange in this reaction, contradictory to the statement made in Ref.~\cite{Hwang2012}.

The claim that a parity spin asymmetry $P_\sigma$ close to unity indicates the dominance of natural parity exchange follows from the following analysis. In the helicity frame, if only $t$-channel meson exchange interactions are present, $P_\sigma$ can be expressed as
\begin{align}
P_\sigma = \dfrac{\sigma^{\rm N} - \sigma^{\rm U}}{\sigma^{\rm N} + \sigma^{\rm U}}, \label{eq:P_sig_N_U}
\end{align}
where $\sigma^{\rm N}$ and $\sigma^{\rm U}$ are the cross sections from natural parity and unnatural parity exchanges, respectively. Thus, if $P_\sigma$ tends to $1$, it means natural parity exchange dominates the reaction. Otherwise, if $P_\sigma$ tends to $-1$, it  means unnatural parity exchange dominates the reaction. The $\kappa$ meson has natural parity while the $K$ meson has unnatural parity. As the LEPS $P_\sigma$ data for $\gamma p\to K^{*0}\Sigma^+$ tends to $1$, in Ref.~\cite{Hwang2012} it is claimed that the $\kappa$ exchange dominates this reaction. However, the analysis based on Eq.~\eqref{eq:P_sig_N_U} is valid only when $t$-channel meson exchange interactions dominate. The LEPS $P_\sigma$ data was measured at $E_\gamma=1.85$--$2.96$ GeV, where $t$-channel meson exchange is not always dominant. In model I of the present work, it is the $s$-channel pole contributions rather than $t$-channel meson exchanges that contribute considerably in this energy and angular region. The approach of $P_\sigma$ to unity results from the interference of various interactions.

Anyway, as illustrated in Fig.~\ref{fig:psig}, the present parity spin asymmetry $P_{\sigma}$ data cannot definitively determine the contributor of the $t$-channel exchanges due to its low measured energy, i.e., $E_{\gamma}=1.85$--$2.96$ GeV. To clarify the role of natural parity exchanges and unnatural parity exchanges based on Eq.~\eqref{eq:P_sig_N_U}, $P_{\sigma}$ should be measured at relatively high energies and at small scattering angles to avoid disturbance from the $s$ and $u$-channel contributions. In Ref.~\cite{newsdme:2023}, the newly released spin density matrix element data for the $\gamma p\to \rho^0 p$ reaction were conducted by the GlueX Collaboration at a photon laboratory energy of $E_{\gamma}= 8.5$ GeV, where contamination from the $s$ and $u$-channel contributions can be avoided, as analyzed in Ref.~\cite{Wang:2025}.

In Fig.~\ref{fig:psig_85}, we show predictions for $P_\sigma$ at $E_\gamma=8.5$ GeV for $\gamma p\to K^{*0}\Sigma^+$ in both models. Here, the black solid lines represent the full results, while the blue dashed, orange dot-double-dashed, and red dotted lines represent the results obtained by switching off the $\Delta(1905)5/2^+$, $\kappa$, and $K$ exchanges, respectively. The upper and lower graphs correspond results in model I and model II, respectively. One sees that in model I, the predicted values of $P_\sigma$ are less than $0.5$, while in model II, the predicted values are close to $1$. The $\kappa$ exchange contributes negligibly in model I while plays a significant role in model II, as the results keep almost unchanged in model I while are reduced significantly in model II when the $\kappa$ exchange is switched off. The future experimental data is expected to distinguish between the two models in our present work and gain a clearer understanding of the reaction mechanism, especially the role of $\kappa$ exchange in this reaction. We mention that at the energy $E_{\gamma}=8.5$ GeV, a Regge type formalism for the reaction amplitudes may be more appropriate than the Feynman type as employed in the present work, which will be carefully investigated in future work.

\section{Summary and conclusion}  \label{sec:summary}

In our previous work \cite{Wang:2018}, we analyzed the differential cross-section data for $\gamma p \to K^{*+} \Sigma^0$ and $\gamma p \to K^{*0} \Sigma^+$ reported by the CLAS Collaboration \cite{Wei:2013, Hleiqawi:2007ad} using an effective Lagrangian approach. It was found that to achieve a satisfactory description of the data, one need to include, in addition to the $t$-channel $K$, $\kappa$, and $K^*$ exchanges, the $s$-channel nucleon ($N$) and $\Delta$ exchanges, and the $u$-channel $\Lambda$, $\Sigma$, $\Sigma^*$ exchanges, at least one $\Delta$ resonance, namely $\Delta(1905)5/2^+$, in the reaction amplitude.

In the present study, we further incorporate data on spin density matrix elements for the $\gamma p \to K^{*0}\Sigma^+$ reaction channel reported by the LEPS Collaboration \cite{Hwang2012} into our analysis. The aim is to impose more stringent constraints on our theoretical models and thus to extract more reliably the reaction mechanism of $K^*\Sigma$ photoproduction reactions, especially to test the role of $\kappa$ exchange in these reactions. In literature, the LEPS data on parity spin asymmetries $P_\sigma$ (which is a combination of spin density matrix elements) for $\gamma p \to K^{*0} \Sigma^+$ at the photon energy $E_\gamma=1.85$--$2.96$ has ever been claimed supporting the dominant $\kappa$ exchange in this reaction as the measured value of $P_\sigma$ approaches unity.

Our results show that with our previously extracted resonance $\Delta(1905)5/2^+$ \cite{Wang:2018} taken into account, the data on both differential cross sections and spin density matrix elements for $\gamma p \to K^{*+} \Sigma^0$ and $\gamma p \to K^{*0} \Sigma^+$ can be simultaneously and satisfactorily described. We obtain two distinct fits, presented as model I and model II in this study, that reproduce the data equally well. In both fits, the $\Delta(1905)5/2^+$ resonance plays an essential role. The two models, however, exhibit a dramatic difference: In model I, the $t$-channel $\kappa$ exchange contributes negligibly, whereas in model II it is significant. This demonstrates that the existing LEPS $P_\sigma$ data, taken at $E_\gamma = 1.85$--$2.96$ GeV, do not mandate a dominant $\kappa$ exchange--contrary to earlier claims \cite{Hwang2012}. The observed $P_\sigma \approx 1$ can also arise from interference involving $s$-channel resonance contributions in the absence of $\kappa$ exchange. 

To resolve this ambiguity, we have presented predictions for the parity spin asymmetry $P_\sigma$ for $\gamma p\to K^{*0}\Sigma^+$ at a much higher energy $E_\gamma=8.5$ GeV at forward angles, where $t$-channel exchanges are expected to dominate. The two models yield clearly distinguishable results: in model I, $P_\sigma < 0.5$, while in model II, $P_\sigma \approx 1$. We therefore conclude that the current low-energy data on spin density matrix elements are insufficient to determine the role of $\kappa$ exchange in $\gamma p \to K^{*0}\Sigma^+$. Future measurements at higher energies, such as those possible at GlueX, may help to discriminate between the two reaction mechanisms identified in this work and to finally settle the question of $\kappa$-exchange dominance. We mention that the predictions for the parity spin asymmetry at $E_\gamma = 8.5$ GeV should be regarded as qualitative rather than quantitative, as the present tree-level effective Lagrangian model is not optimized for this high-energy region. Therefore, these predictions carry a systematic uncertainty associated with the limitations of the model.

\begin{acknowledgments}
This work is partially supported by the National Natural Science Foundation of China under Grants No.~12575093, No.~12305097, No.~12305137, and No.~12175240, and the Fundamental Research Funds for the Central Universities. A.C.W. is supported by the Shandong Provincial Natural Science Foundation, China (Grants No.~ZR2024QA096 and No.~ZR2026MS0033), and the Taishan Scholar Young Talent Program (Grant No.~tsqn202408091).
\end{acknowledgments}

\end{document}